\pdfoutput=1
\documentclass[12pt,a4paper]{article}
\usepackage{ifthen} % for conditional statements
\newboolean{pdflatex}
\setboolean{pdflatex}{true} % False for eps figures 

\newboolean{articletitles}
\setboolean{articletitles}{true} % False removes titles in references

\newboolean{uprightparticles}
\setboolean{uprightparticles}{false} %True for upright particle symbols

\def\paperauthors{LHCb collaboration} % Leave as is for PAPER, CONF and FIGURE
\def\paperasciititle{B-jet fragmentation with charged B-meson decays in 13 TeV pp collisions at LHCb} % Set ASCII title here !! MAKE sure it's only ASCII characters !! 
\def\papertitle{\B-jet fragmentation with $\Bpm \to \jpsi \Kpm$ decays in $\sqs = 13\tev$ $pp$ collisions at LHCb} % Latex formatted title
\def\paperkeywords{{High Energy Physics}, {LHCb}} % Comma separated list
\def\papercopyright{\the\year\ CERN for the benefit of the LHCb collaboration} % new since 9/Apr/2018
\def\paperlicence{CC BY 4.0 licence}
\def\paperlicenceurl{https://creativecommons.org/licenses/by/4.0/}

\newif\ifEnableSectionTOCLinks
\EnableSectionTOCLinksfalse % deactivated

\usepackage[top=1in, bottom=1.25in, left=1in, right=1in]{geometry}

\usepackage{microtype}
\usepackage{lineno}  % for line numbering during review
\usepackage{xspace} % To avoid problems with missing or double spaces after
\usepackage{caption} %these three command get the figure and table captions automatically small

\usepackage{graphicx}  % to include figures (can also use other packages)
\usepackage{color}
\usepackage{colortbl}
\graphicspath{{./figs/}} % Make Latex search fig subdir for figures
\usepackage{amsmath} % Adds a large collection of math symbols
\usepackage{amssymb}
\usepackage{amsfonts}
\usepackage{upgreek} % Adds in support for greek letters in roman typeset

\newcommand*\patchAmsMathEnvironmentForLineno[1]{%
\expandafter\let\csname old#1\expandafter\endcsname\csname #1\endcsname
\expandafter\let\csname oldend#1\expandafter\endcsname\csname
end#1\endcsname
 \renewenvironment{#1}%
   {\linenomath\csname old#1\endcsname}%
   {\csname oldend#1\endcsname\endlinenomath}%
}
\newcommand*\patchBothAmsMathEnvironmentsForLineno[1]{%
  \patchAmsMathEnvironmentForLineno{#1}%
  \patchAmsMathEnvironmentForLineno{#1*}%
}
\AtBeginDocument{%
\patchBothAmsMathEnvironmentsForLineno{equation}%
\patchBothAmsMathEnvironmentsForLineno{align}%
\patchBothAmsMathEnvironmentsForLineno{flalign}%
\patchBothAmsMathEnvironmentsForLineno{alignat}%
\patchBothAmsMathEnvironmentsForLineno{gather}%
\patchBothAmsMathEnvironmentsForLineno{multline}%
\patchBothAmsMathEnvironmentsForLineno{eqnarray}%
}

\usepackage[pdftex,
            pdfauthor={\paperauthors},
            pdftitle={\paperasciititle},
            pdfkeywords={\paperkeywords}]{hyperref}
\usepackage{hyperxmp}
\hypersetup{
    pdfcopyright={Copyright (C) \papercopyright},
    pdflicenseurl={\paperlicenceurl}
}
\usepackage[colorinlistoftodos,textsize=scriptsize]{todonotes}

\usepackage[bottom,flushmargin,hang,multiple]{footmisc}

\usepackage[all]{hypcap} % Internal hyperlinks to floats.

\usepackage{xspace} 
\usepackage{upgreek}

\def\lhcb   {\mbox{LHCb}\xspace}

\def\aleph  {\mbox{ALEPH}\xspace}
\def\delphi {\mbox{DELPHI}\xspace}
\def\opal   {\mbox{OPAL}\xspace}

\def\sld    {\mbox{SLD}\xspace}
\def\MagUp {\mbox{\em Mag\kern -0.05em Up}\xspace}

\def\hltone {HLT1\xspace}
\def\hlttwo {HLT2\xspace}

\ifthenelse{\boolean{uprightparticles}}%
{

 \def\Pmu         {\ensuremath{\upmu}\xspace}

 \def\Ppi         {\ensuremath{\uppi}\xspace}

 \def\Pchi        {\ensuremath{\upchi}\xspace}                 
 \def\Ppsi        {\ensuremath{\uppsi}\xspace}

 \def\PDelta      {\ensuremath{\Delta}\xspace}                 
 \def\PXi         {\ensuremath{\Xi}\xspace}                 
 \def\PLambda     {\ensuremath{\Lambda}\xspace}                 
 \def\PSigma      {\ensuremath{\Sigma}\xspace}                 
 \def\POmega      {\ensuremath{\Omega}\xspace}                 
 \def\PUpsilon    {\ensuremath{\Upsilon}\xspace}
 \let\oldPi\Pi
 \def\PPi         {\ensuremath{\oldPi}\xspace}

 \def\PB      {\ensuremath{\mathrm{B}}\xspace}                 
 \def\PD      {\ensuremath{\mathrm{D}}\xspace}                 
 \def\PJ      {\ensuremath{\mathrm{J}}\xspace}                 
 \def\PK      {\ensuremath{\mathrm{K}}\xspace}                 
 \def\PZ      {\ensuremath{\mathrm{Z}}\xspace}                 
 \def\Pb      {\ensuremath{\mathrm{b}}\xspace}                 
 \def\Pc      {\ensuremath{\mathrm{c}}\xspace}                 
                  
 \def\Pe      {\ensuremath{\mathrm{e}}\xspace}                 
 \def\Pp      {\ensuremath{\mathrm{p}}\xspace}                 

 \def\Ps      {\ensuremath{\mathrm{s}}\xspace}

 \def\thebaroffset{0.0em}
}
{

 \def\Pmu         {\ensuremath{\mu}\xspace}

 \def\Ppi         {\ensuremath{\pi}\xspace}

 \def\Pchi        {\ensuremath{\chi}\xspace}                 
 \def\Ppsi        {\ensuremath{\psi}\xspace}                 
                  
 \mathchardef\PDelta="7101
 \mathchardef\PXi="7104
 \mathchardef\PLambda="7103
 \mathchardef\PSigma="7106
 \mathchardef\POmega="710A
 \mathchardef\PUpsilon="7107
 \mathchardef\PPi="7105
 \def\PB      {\ensuremath{B}\xspace}                 
 \def\PD      {\ensuremath{D}\xspace}                 
 \def\PJ      {\ensuremath{J}\xspace}                 
 \def\PK      {\ensuremath{K}\xspace}                 
 \def\PZ      {\ensuremath{Z}\xspace}                 
 \def\Pb      {\ensuremath{b}\xspace}                 
 \def\Pc      {\ensuremath{c}\xspace}                 
                  
 \def\Pe      {\ensuremath{e}\xspace}                 
 \def\Pp      {\ensuremath{p}\xspace}                 

 \def\Ps      {\ensuremath{s}\xspace}

 \def\thebaroffset{0.18em}
}
\newcommand{\offsetoverline}[2][\thebaroffset]{\kern #1\overline{\kern -#1 #2}}%

\makeatletter
\ifcase \@ptsize \relax% 10pt
  \newcommand{\miniscule}{\@setfontsize\miniscule{4}{5}}% \tiny: 5/6
\or% 11pt
  \newcommand{\miniscule}{\@setfontsize\miniscule{5}{6}}% \tiny: 6/7
\or% 12pt
  \newcommand{\miniscule}{\@setfontsize\miniscule{5}{6}}% \tiny: 6/7
\fi
\makeatother

\DeclareRobustCommand{\optbar}[1]{\shortstack{{\miniscule (\rule[.5ex]{1.25em}{.18mm})}
  \\ [-.7ex] $#1$}}

\def\epem       {{\ensuremath{\Pe^+\Pe^-}}\xspace}

\def\mup        {{\ensuremath{\Pmu^+}}\xspace}
\def\mun        {{\ensuremath{\Pmu^-}}\xspace} % muon negative (\mum is taken)

\def\Z      {{\ensuremath{\PZ}}\xspace}

\def\squark    {{\ensuremath{\Ps}}\xspace}

\def\cquark    {{\ensuremath{\Pc}}\xspace}

\def\bquark    {{\ensuremath{\Pb}}\xspace}

\def\pion   {{\ensuremath{\Ppi}}\xspace}

\def\pipm   {{\ensuremath{\pion^\pm}}\xspace}

\def\kaon    {{\ensuremath{\PK}}\xspace}
\def\KorKbar {\kern \thebaroffset\optbar{\kern -\thebaroffset \PK}{}\xspace}

\def\Kpm     {{\ensuremath{\kaon^\pm}}\xspace}

\def\D       {{\ensuremath{\PD}}\xspace}

\def\DorDbar {\kern \thebaroffset\optbar{\kern -\thebaroffset \PD}\xspace}

\def\Dp      {{\ensuremath{\D^+}}\xspace}
\def\Dm      {{\ensuremath{\D^-}}\xspace}

\def\DpDm    {\ensuremath{\Dp {\kern -0.16em \Dm}}\xspace}

\def\B       {{\ensuremath{\PB}}\xspace}

\def\BorBbar {\kern \thebaroffset\optbar{\kern -\thebaroffset \PB}\xspace}

\def\Bd      {{\ensuremath{\B^0}}\xspace}

\def\BdorBdbar {\kern \thebaroffset\optbar{\kern -\thebaroffset \Bd}\xspace}

\def\Bpm     {{\ensuremath{\B^\pm}}\xspace}

\def\Bs      {{\ensuremath{\B^0_\squark}}\xspace}

\def\BsorBsbar {\kern \thebaroffset\optbar{\kern -\thebaroffset \Bs}\xspace}

\def\jpsi     {{\ensuremath{{\PJ\mskip -3mu/\mskip -2mu\Ppsi}}}\xspace}
\def\psitwos  {{\ensuremath{\Ppsi{(2S)}}}\xspace}

\def\chicone  {{\ensuremath{\Pchi_{\cquark 1}}}\xspace}

\def\Y#1S{\ensuremath{\PUpsilon{(#1S)}}\xspace}

\def\theX     {{\ensuremath{\Pchi_{c1}(3872)}}\xspace}

\def\proton      {{\ensuremath{\Pp}}\xspace}

\def\Lz          {{\ensuremath{\PLambda}}\xspace}

\def\LorLbar     {\kern \thebaroffset\optbar{\kern -\thebaroffset \PLambda}\xspace}

\def\Lc          {{\ensuremath{\Lz^+_\cquark}}\xspace}

\newcommand{\decay}[2]{\ensuremath{\mathinner{#1\!\to #2}}\xspace}

\def\to                 {\ensuremath{\rightarrow}\xspace}

\def\AT#1     {\ensuremath{A_{\mathrm{T}}^{#1}}\xspace}           % 2

\def\C#1      {\ensuremath{\mathcal{C}_{#1}}\xspace}                       % 9
\def\Cp#1     {\ensuremath{\mathcal{C}_{#1}^{'}}\xspace}                    % 7
\def\Ceff#1   {\ensuremath{\mathcal{C}_{#1}^{\mathrm{(eff)}}}\xspace}        % 9  
\def\Cpeff#1  {\ensuremath{\mathcal{C}_{#1}^{'\mathrm{(eff)}}}\xspace}       % 7
\def\Ope#1    {\ensuremath{\mathcal{O}_{#1}}\xspace}                       % 2
\def\Opep#1   {\ensuremath{\mathcal{O}_{#1}^{'}}\xspace}                    % 7

\newcommand{\nospaceunit}[1]{\ensuremath{\text{#1}}}       
\newcommand{\aunit}[1]{\ensuremath{\text{\,#1}}}       
\newcommand{\tev}{\aunit{Te\kern -0.1em V}\xspace}
\newcommand{\gev}{\aunit{Ge\kern -0.1em V}\xspace}
\newcommand{\mev}{\aunit{Me\kern -0.1em V}\xspace}
\newcommand{\kev}{\aunit{ke\kern -0.1em V}\xspace}
\newcommand{\ev}{\aunit{e\kern -0.1em V}\xspace}
 
\newcommand{\mevc}{\ensuremath{\aunit{Me\kern -0.1em V\!/}c}\xspace}
\newcommand{\gevc}{\ensuremath{\aunit{Ge\kern -0.1em V\!/}c}\xspace}
\newcommand{\mevcc}{\ensuremath{\aunit{Me\kern -0.1em V\!/}c^2}\xspace}
\newcommand{\gevcc}{\ensuremath{\aunit{Ge\kern -0.1em V\!/}c^2}\xspace}
\def\mum  {\ensuremath{\,\upmu\nospaceunit{m}}\xspace}

\def\fb   {\ensuremath{\aunit{fb}}\xspace}
\def\invfb   {\ensuremath{\fb^{-1}}\xspace}

\def\deriv {\ensuremath{\mathrm{d}}}

\def\gsim{{~\raise.15em\hbox{$>$}\kern-.85em
          \lower.35em\hbox{$\sim$}~}\xspace}
\def\lsim{{~\raise.15em\hbox{$<$}\kern-.85em
          \lower.35em\hbox{$\sim$}~}\xspace}

\def\sqs   {\ensuremath{\protect\sqrt{s}}\xspace}

\def\pt         {\ensuremath{p_{\mathrm{T}}}\xspace}

\def\ptot       {\ensuremath{p}\xspace}

\def\evtgen     {\mbox{\textsc{EvtGen}}\xspace}

\def\geant      {\mbox{\textsc{Geant4}}\xspace}

\def\photos     {\mbox{\textsc{Photos}}\xspace}

\def\pythia     {\mbox{\textsc{Pythia}}\xspace}

\def\tell1  {TELL1\xspace}
\def\ukl1   {UKL1\xspace}

\newcommand{\eg}{\mbox{\itshape e.g.}\xspace}
\newcommand{\ie}{\mbox{\itshape i.e.}\xspace}

\newcommand{\lhcborcid}[1]{\href{https://orcid.org/#1}{\hspace*{0.1em}\raisebox{-0.45ex}{\includegraphics[width=1em]{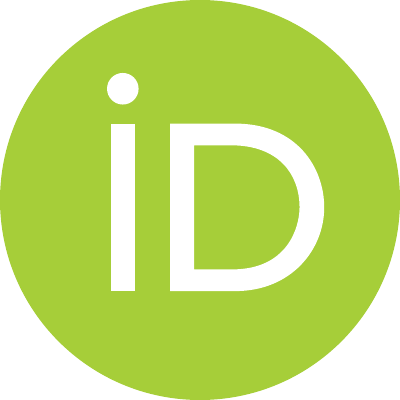}}}}

\hypersetup{
  colorlinks   = true, %Colours links instead of ugly boxes
  urlcolor     = blue, %Colour for external hyperlinks
  linkcolor    = blue, %Colour of internal links
  citecolor    = red   %Colour of citations
}

\ifEnableSectionTOCLinks
    \usepackage[explicit]{titlesec} % to change headings
    
    \let\oldcontentsline\contentsline
    \renewcommand

    \titleformat{\section}{\normalfont\Large\bf}{\hyperlink{tocsection.\thesection}{{\thesection} \parbox[t]{\dimexpr\textwidth-1pc}{#1}}}{1pc}{}

    \titleformat{\subsection}{\normalfont\bf}{\hyperlink{tocsubsection.\thesubsection}{{\thesubsection} \parbox[t]{\dimexpr\textwidth-1pc}{#1}}}{1pc}{}

    \titleformat{name=\section,numberless}[display]{}{}{0pt}{\normalfont\Huge\bfseries #1}
\fi

\usepackage{cite} % Allows for ranges in citations
\usepackage{LHCb/mciteplus}
\usepackage{longtable} % only for template; not usually to be used in PAPERs
\usepackage{float}

\def\jt         {\ensuremath{j_{\mathrm{T}}}\xspace}
\def\ptj         {\ensuremath{p_{\mathrm{T,jet}}}\xspace}
\def\ptB         {\ensuremath{p_{\mathrm{T,}B^{\pm}}}\xspace}

\begin{document}

%%%%%%%%%%%%%%%%%%%%%%%%%
%%%%% Title     %%%%%%%%%
%%%%%%%%%%%%%%%%%%%%%%%%%
\renewcommand{\thefootnote}{\fnsymbol{footnote}}
\setcounter{footnote}{1}

% %%%%%%% CHOOSE TITLE PAGE--------
%\onecolumn
%\input{title-LHCb-PAPER}
%\twocolumn
% %%%%%%%%%%%%% ---------

%%%%%%%%%%%%%%%%%%%%%%%%%
%%%%%  TITLE PAGE  %%%%%%
%%%%%%%%%%%%%%%%%%%%%%%%%
\begin{titlepage}
\pagenumbering{roman}

% Header ---------------------------------------------------
\vspace*{-1.5cm}
\centerline{\large EUROPEAN ORGANIZATION FOR NUCLEAR RESEARCH (CERN)}
\vspace*{1.5cm}
\noindent
\begin{tabular*}{\linewidth}{lc@{\extracolsep{\fill}}r@{\extracolsep{0pt}}}
\ifthenelse{\boolean{pdflatex}}% Logo format choice
{\vspace*{-1.5cm}\mbox{\!\!\!\includegraphics[width=.14\textwidth]{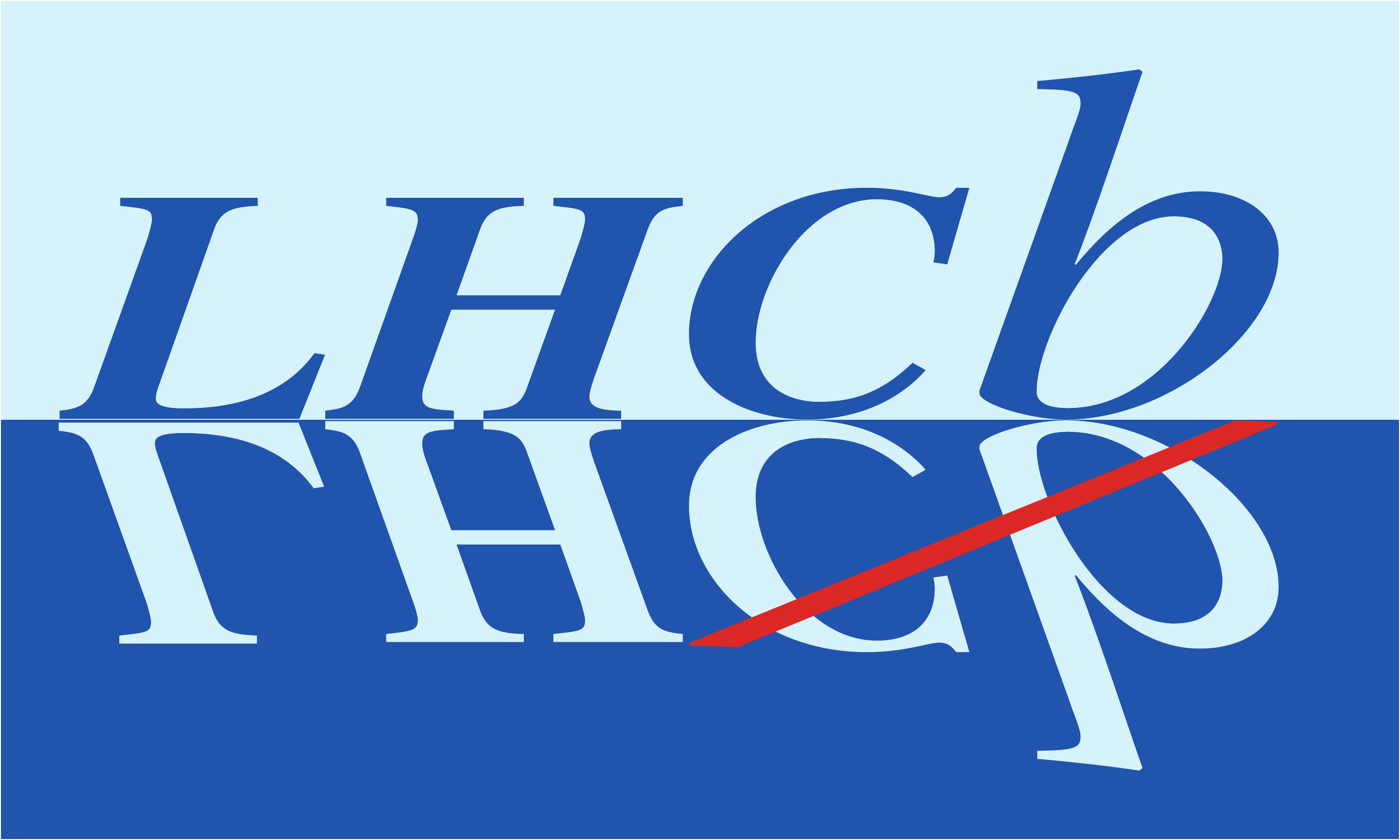}} & &}%
{\vspace*{-1.2cm}\mbox{\!\!\!\includegraphics[width=.12\textwidth]{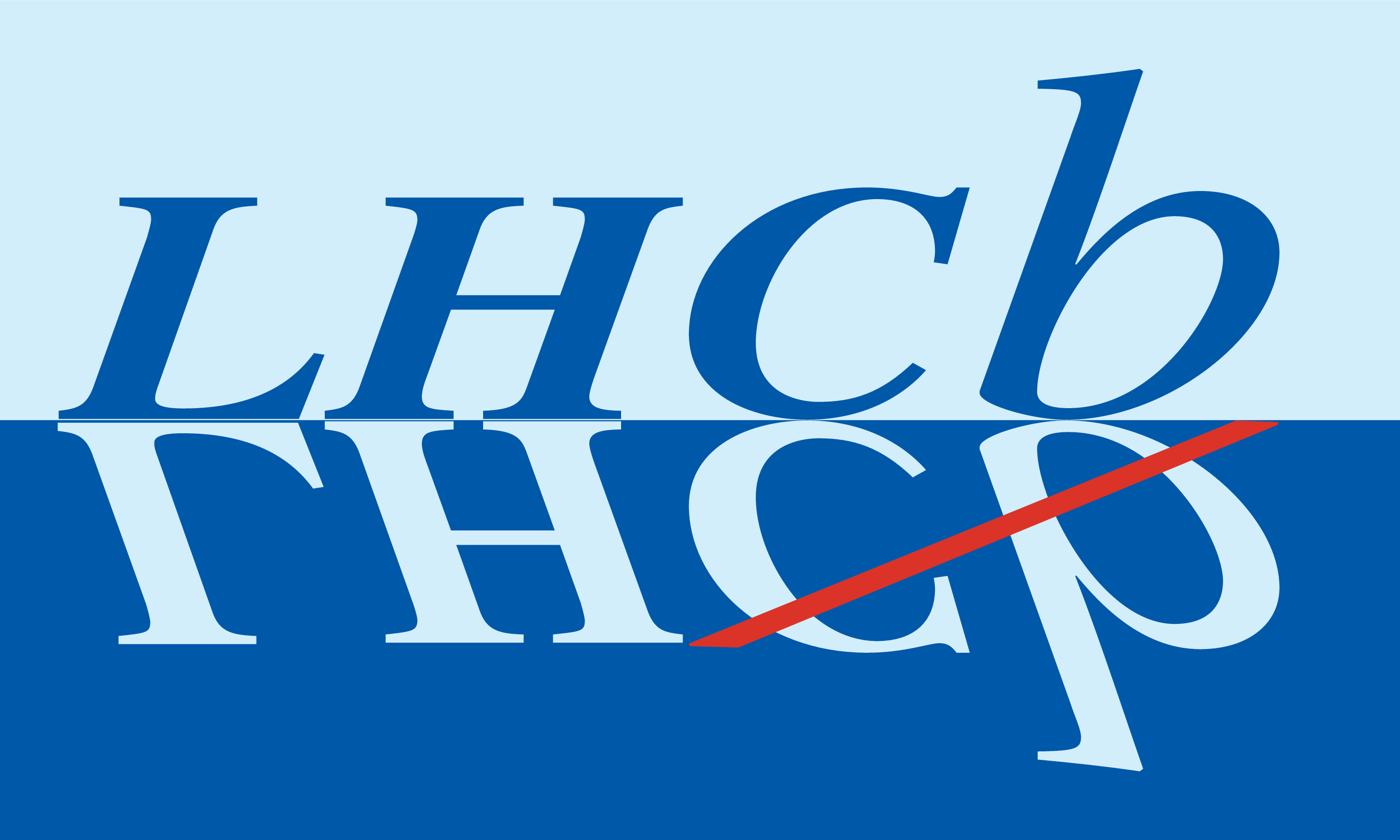}} & &}%
\\
 & & \today \\ % Date - Can also hardwire e.g.: 23 March 2010
 & & \\
% not in paper \hline
\end{tabular*}

\vspace*{4.0cm}

% Title --------------------------------------------------
{\normalfont\bfseries\boldmath\huge
\begin{center}
% DO NOT EDIT HERE. Instead edit macro in main.tex to keep metadata correct
  \papertitle 
\end{center}
}

\vspace*{2.0cm}

% Authors -------------------------------------------------
\begin{center}
%In the footnote, replace 'paper' by 'Letter' in case of submission to PRL or PLB 
% Edit macro in main.tex to keep metadata correct
\paperauthors\footnote{Authors are listed at the end of this paper.}
\end{center}

\vspace{\fill}

% Abstract -----------------------------------------------
\begin{abstract}
  \noindent  
  The collinear and transverse-momentum-dependent jet fragmentation function and the radial profile for \Bpm mesons in jets are measured. The \Bpm mesons are reconstructed through the $\jpsi (\to \mup \mun) \Kpm$ decay channel using proton-proton collision data collected during 2016--2018 with the LHCb detector at a center-of-mass (c.m.) energy of $\sqs=13\tev$, corresponding to an integrated luminosity of $5.4\invfb$. The results complement recent measurements of jet fragmentation functions for heavy-flavor hadrons and suggest a growing contribution of gluon fragmentation to \Bpm mesons as the jet transverse momentum increases.
  
\end{abstract}

\vspace*{2.0cm}

\begin{center}
  Published in Physical Review D113 (2026) 112006 
\end{center}

\vspace{\fill}

{\footnotesize 
% Edit macro in main.tex to keep metadata correct
\centerline{\copyright~\papercopyright. \href{\paperlicenceurl}{\paperlicence}.}}
\vspace*{2mm}

\end{titlepage}

\renewcommand{\thefootnote}{\arabic{footnote}}
\setcounter{footnote}{0}

%%%%%%%%%%%%%%%%%%%%%%%%%%%%%%%%
%%%%%  Table of Content   %%%%%%
%%%%%%%%%%%%%%%%%%%%%%%%%%%%%%%%
%%%% Uncomment if desired
%\tableofcontents

\cleardoublepage

%%%%%%%%%%%%%%%%%%%%%%%%%
%%%%% Main text %%%%%%%%%
%%%%%%%%%%%%%%%%%%%%%%%%%

\pagestyle{plain} % restore page numbers for the main text
\setcounter{page}{1}
\pagenumbering{arabic}

%% Uncomment during review phase. 
%% Comment before a final submission.
%\linenumbers

%% This is the main body
%% It is useful to have a single file so comments are not missed in overleaf.
%\input{body}

%%% Adding body to main.tex for PRD submission %%%
\section{Introduction}
\label{sec:Introduction}
       The processes by which color-charged quarks and gluons form color-neutral bound states, \ie~hadronize, are still poorly understood. Given that the full transition from partons to hadrons includes a nonperturbative regime, functions describing hadronization must be constrained through measurements of final-state hadrons. From such measurements, collinear fragmentation functions (FFs) are extracted, representing the probability that an outgoing parton will produce a particular species of hadron with a given collinear momentum fraction with respect to the parent parton. Transverse-momentum-dependent (TMD) FFs that additionally describe dependence on hadron momentum transverse to the outgoing parton momentum direction are also extracted~\cite{parton_FFs}. 

       While FFs are well constrained for many light quark species (see \eg Refs.~\cite{light_charged_FFs, light_neutral_FFs}), their heavy quark counterparts have received far less attention, and, in particular for beauty hadrons, few measurements exist~\cite{ALEPH_Bfrag, OPAL_Bfrag, SLD_Bfrag, DELPHI_Bfrag, ATLAS_Bfrag, ATLAS_partialBfrag, CMS_partialBfrag}. In calculations of processes involving heavy-flavor hadrons in the final state, the heavy-quark mass $m_Q$ constitutes an additional, non-negligible scale, along with the hard-interaction scale $Q$, where the jet transverse momentum \pt is often used as a proxy ~\cite{DstarFF_pheno}. These multiscale processes serve as a stringent test of Quantum Chromodynamics (QCD) and the applicability of various factorization theorems. Additionally, improved measurements and modeling of beauty fragmentation may for example aid in the search for di-Higgs production in channels involving $b$ quarks.
       
       Measurements of $B$-hadron fragmentation from the \aleph, \opal, \delphi, and \sld experiments were performed in  \epem collision systems and have been used to constrain collinear FFs~\cite{ALEPH_Bfrag, OPAL_Bfrag, SLD_Bfrag, DELPHI_Bfrag}. These data suggest that heavy-flavor hadrons carry a significantly larger collinear momentum fraction than their light-flavor counterparts. Hadron production in \epem collisions serves as a primary input for global analyses of FFs due to the absence of hadrons in the initial state. However, gluon fragmentation functions are poorly constrained from these data due to the lack of access to gluons at leading order. This highlights the usefulness of complementary measurements in hadron-hadron collisions, where gluons are directly accessible at leading order in $\alpha_{s}$, allowing for the contribution of gluon fragmentation to heavy-flavor hadrons to be studied with higher precision. In \epem collisions, the collinear momentum fraction is determined as the fraction of energy carried by the beauty hadron with respect to half of the c.m. energy. The situation in proton-proton ($pp$) collisions is more complicated, as protons are composite and the beam energy cannot be used to reliably measure the outgoing parton energy. Instead, reconstructed jets are often used as a proxy for the outgoing parton at leading order. In this case, jet fragmentation functions (JFFs) can be measured, defined as the cross section of hadron-in-jet production normalized by the jet production cross section~\cite{JFF_SCET}. In recent global analyses~\cite{DstarFF_pheno,light_charged_FFs, light_neutral_FFs}, JFFs have been included in a consistent framework with single-hadron production measurements in \epem and $pp$ to extract information on the relevant FFs.
        
        Many methods have been proposed to study the fragmentation and hadronization process by exploring the substructure of jets (see \eg Refs.~\cite{JFF_SCET, JFF_Zjets, LJP, EEC_overview, beauty_EECs} or Ref.~\cite{jet_substructure_overview} for a recent review). One approach reconstructs subjets from jet constituents to extract information on the parton shower evolution, while another is focused on how hadrons are distributed within jets and correlated with the jet axis, to extract information on the bound-state formation. Both are necessary for a full picture of parton-to-jet evolution, as the radiation pattern of color charges during the parton shower can affect the distribution of hadrons in the final state. For heavy quarks in particular, the dead-cone effect was predicted, defined by the suppression of radiation in a conical region around the radiating quark and characterized by the dead-cone angle, $\theta_{q} = m_{q}/E_{q}$~\cite{deadcone_theory}. It was measured for the first time for jets with charm hadrons by the ALICE Collaboration~\cite{ALICE_deadcone}. It was observed again in an analysis of OPAL and DELPHI data for $b$- and $c$-quark jets~\cite{LEP_deadcone}. Recent measurements in the beauty sector from LHCb also show strong evidence for the effect~\cite{LHCb-PAPER-2025-010, LHCb-PAPER-2025-038}. All of these analyses use different methods but point towards the dead-cone effect. In addition, the jet mass has been measured for the first time in \B-tagged jets by the LHCb Collaboration~\cite{LHCb-PAPER-2025-009}, which gives a complementary handle on how these quantities depend on the virtuality of the initiating parton, where highly virtual partons could lead to significantly different radiation patterns and final-state hadron distributions. 
        
\begin{figure}[t]		
            \centering
            \includegraphics[scale=1.0]{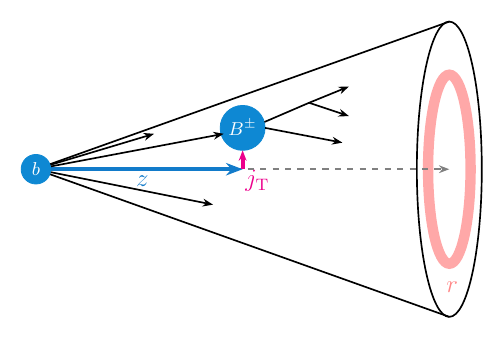}
            \caption{Schematic of the collinear momentum fraction $z$, transverse momentum with respect to the jet axis \jt, and radial profile $r$ of a $\Bpm$ meson within a jet.}
            \label{fig:Bjet_observables}
        \end{figure}

        The focus of this paper is on measuring correlations between jets and their constituent \Bpm mesons in their decays to {\jpsi \decay{(}{\mup \mun)} \Kpm}. In particular, jet fragmentation functions are measured through the collinear momentum fraction 
        
        \begin{equation}
            z \equiv \frac{\vec{p}_{\text{\Bpm}} \cdot \vec{p}_{\text{jet}}}{| \vec{p}_{\text{jet}}|^{2}}
            \label{eq:frag_z},
        \end{equation}
        defined as the projection of the \Bpm momentum vector on the jet momentum vector, and the transverse momentum with respect to the jet axis
        \begin{equation}
            \jt \equiv \frac{|\vec{p}_{\text{\Bpm}} \times \vec{p}_{\text{jet}}|}{| \vec{p}_{\text{jet}}|}.
            \label{eq:frag_jt}
        \end{equation} 
        Collinear and TMD JFFs, denoted as $F(z)$ and $F(z, \jt)$, respectively, provide useful information on hadron-in-jet fragmentation. The radial profile, which offers additional insights into the jet fragmentation and hadronization process, is defined as
        \begin{equation}
            r \equiv \Delta R(\text{\Bpm}, \text{jet}) = \sqrt{(y_{\text{\Bpm}} - y_{\text{jet}})^{2} + (\phi_{\text{\Bpm}} - \phi_{\text{jet}})^{2} }.
            \label{eq:frag_r}
        \end{equation}   
        The variables $\vec{p}$, $y$, and $\phi$ represent the three-momentum, rapidity, and azimuthal angle (measured in the laboratory frame), respectively. In the leading-order picture, in the limit that the jet is a perfect proxy for the hard-scattered parton, $z$ and \jt correspond exactly to variables appearing in collinear and TMD FFs. In practice, there are more subtleties in interpreting the experimental data through the measured JFFs (see Refs.~\cite{JFF_SCET, JFF_Zjets} for details). A schematic of the observables described by Eqs.~\ref{eq:frag_z}--\ref{eq:frag_r} is shown in Fig.~\ref{fig:Bjet_observables}.
        
        The collinear JFF has been measured for charged hadrons in the case of inclusive jets by the ATLAS Collaboration~\cite{ATLAS_frag}, and for $Z$-tagged~\cite{LHCB-PAPER-2019-012, LHCB-PAPER-2022-013} and heavy-flavor-tagged~\cite{LHCb-PAPER-2025-038} jets by the LHCb Collaboration. These data provide constraints on the hadronization process for gluon, light-quark, and heavy-flavor-quark-initiated jets, respectively. For the $Z$-tagged jets from the LHCb experiment, there also exists a measurement of TMD JFFs with well-identified charged pions, charged kaons, and protons, providing sensitivity to flavor-dependent TMD FFs~\cite{LHCB-PAPER-2022-013}. In the heavy-quark sector, measurements of collinear JFFs for fully reconstructed hadrons in jets exist with \jpsi~\cite{LHCB-PAPER-2016-064}, $\psi(2S)$ and $\chicone(3872)$ states~\cite{LHCb-PAPER-2024-021} from the LHCb Collaboration, $\Bpm$ mesons~\cite{ATLAS_Bfrag} and $D^{*+}$ mesons~\cite{ATLAS_Dstarfrag} from the ATLAS Collaboration, and $D^{0}$ mesons~\cite{ALICE_D0frag_1, ALICE_D0frag_2} and \Lc baryons~\cite{ALICE_Lcfrag} from the ALICE experiment. There also exist similar measurements for collinear JFFs for partially reconstructed inclusive \B-hadron decays by the ATLAS~\cite{ATLAS_partialBfrag} and CMS~\cite{CMS_partialBfrag} collaborations. The measurements presented in this paper, performed in $pp$ collisions at a c.m. energy \sqs = 13\tev, complement these experimental efforts as well as the previous measurements constraining $B$ fragmentation functions. The collinear JFF along with 1D distributions of \jt, and $r$ are reported. In addition, the TMD JFF in $(z, \jt)$ as well as the joint distributions of $(z, r)$ and $(\jt, r)$ are measured for the first time for \Bpm hadrons.

\section{Detector and simulation framework}
\label{sec:detector}

The \lhcb detector~\cite{LHCb-DP-2008-001,LHCb-DP-2014-002} is a single-arm forward spectrometer covering the \mbox{pseudorapidity} range $2<\eta <5$, designed for the study of particles containing \bquark or \cquark
quarks. The detector used to collect the data analyzed in this paper includes a high-precision tracking system consisting of a silicon-strip vertex detector surrounding the $pp$ interaction region~\cite{LHCb-DP-2014-001}, a large-area silicon-strip detector located upstream of a dipole magnet with a bending power of about $4{\mathrm{\,T\,m}}$, and three stations of silicon-strip detectors and straw drift tubes~\cite{LHCb-DP-2017-001} placed downstream of the magnet. The tracking system provides a measurement of the momentum, \ptot, of charged particles with a relative uncertainty that varies from 0.5\% at low momentum to 1.0\% at 200\gevc. The minimum distance of a track to a primary $pp$ collision vertex (PV), the impact parameter, is measured with a resolution of $(15+29/\pt)\mum$, where \pt is the component of the momentum transverse to the beam, in\,\gevc. Different types of charged hadrons are distinguished using information from two ring-imaging Cherenkov (RICH) detectors~\cite{LHCb-DP-2012-003}. Photons, electrons and hadrons are identified by a calorimeter system consisting of scintillating-pad and preshower detectors, an electromagnetic and a hadronic calorimeter. Muons are identified by a system composed of alternating layers of iron and multiwire proportional chambers~\cite{LHCb-DP-2012-002}.

The online event selection is performed by a trigger~\cite{LHCb-DP-2012-004}, which consists of a hardware stage, based on information from the calorimeter and muon systems, followed by a software stage, which applies a full event reconstruction. In particular, several online selection requirements are imposed on the muons and \jpsi candidate, starting with hardware trigger requirements on the geometric mean \mbox{$\sqrt{\pt(\mup)\pt(\mun)} > (1.3-1.5) \gevc$} (where the threshold varied during the data taking period) and on the number of hits in the scintillating-pad detector $N_{\rm SPD} < 900$. Further selection requirements are imposed at two subsequent software trigger stages (\hltone/\hlttwo) on the muon momentum, muon particle identification  (PID), track and vertex quality, as well as the dimuon invariant mass, to identify $\mu^{+}\mu^{-}$ pairs consistent with a \jpsi meson, displaced with respect to the PV.

Simulation is required to model the effects of the detector acceptance and the imposed selection requirements. In the simulation, $pp$ collisions are generated using \pythia~\cite{Sjostrand:2007gs} with a specific \lhcb configuration~\cite{LHCb-PROC-2010-056}. Decays of unstable particles are described by \evtgen~\cite{Lange:2001uf}, in which final-state radiation is generated using \photos~\cite{davidson2015photos}. The interaction of the generated particles with the detector, and its response, are implemented using the \geant toolkit~\cite{Allison:2006ve, Agostinelli:2002hh} as described in Ref.~\cite{LHCb-PROC-2011-006}. The outcome of the simulation chain yields signal candidates at both the generator, or ``truth'' level, as well as at the ``reconstructed'' level (after being processed through the detector simulation), from which the detector performance can be studied in detail. The response of the RICH detectors in the detector simulation is calibrated using \decay{D^{*+}}{D^{0}\pi^{+}} and \decay{D^{0}}{K^{-}\pi^{+}} decays for charged kaon and pion PID and \decay{\Lc}{pK^{-}\pi^{+}} decays for proton PID~\cite{LHCb-DP-2018-001}.\footnote{This includes the charge conjugate decay in both cases.}

\section{Analysis strategy}
\label{sec:methods}

    \subsection{Event and candidate selection}

        Only events with exactly only one reconstructed PV are used in the analysis to allow for a more precise determination of the jet kinematics. The \Bpm mesons are reconstructed using the \decay{\Bpm} {\jpsi \decay{(}{\mup \mun)} \Kpm} decay channel. Well-identified pairs of oppositely charged muons, each with transverse momentum $\pt > 500\,\mevc$, originating from a common vertex are combined to form \jpsi candidates. The invariant mass of the pair is required to lie within 100 \mevcc of the known \jpsi mass, and the \jpsi candidate must be significantly displaced from the PV such that it is consistent with originating from the decay of a \bquark hadron. The \jpsi candidates are combined with a well-identified kaon which must have $\pt>250\mevc$, a good quality track fit, and be displaced from the PV, to form a \Bpm candidate. The \Bpm candidates are required to have an invariant mass in the range $5.15 < m_{\mu\mu K} < 5.55\gevcc$.
        
        Once the $\Bpm$ candidate is reconstructed, it is provided to the jet clustering algorithm (replacing its decay products), along with other final-state particles as determined by a particle flow algorithm~\cite{LHCb-PAPER-2013-058}. Jets are reconstructed using \texttt{FastJet 3.4.1} software~\cite{fastjet} with the anti-$k_{t}$ algorithm and a resolution parameter $R=0.5$ (AK5)~\cite{Cacciari:2008gp}. The jet axis is determined with the E-scheme, where the four-momentum of the jet is defined as the sum of the four-momentum of its constituents. Jets are required to contain a reconstructed \Bpm candidate, and have $2.5 < y_{\rm jet} < 4.0$ to ensure all of the constituents are within the LHCb acceptance. The results are reported within the range $ 10 < \ptj < 100$ \gevc in six different bins.

    \subsection{Signal extraction}
    \label{sec:signal_extraction}
        
\begin{figure}[tb]		
            \centering
            \includegraphics[scale=0.6]{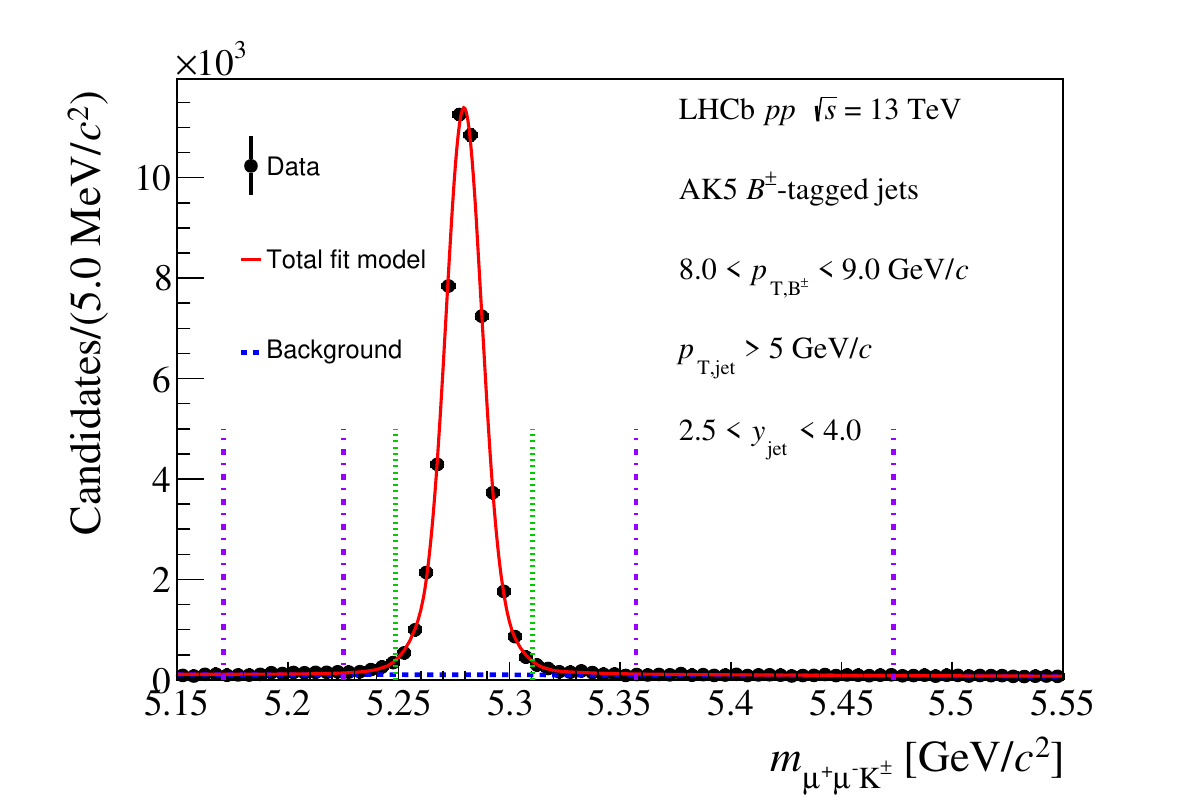}
            \caption{Example of the invariant mass distribution of the \decay{\Bpm} {\jpsi \decay{(}{\mup \mun)} \Kpm} candidates within $8 < \ptB < 9$ \gevc. The data points are shown as black circles while the total fit model is shown as a red curve. The blue dashed line depicts the combinatorial background model, while the green dotted and purple dash-dotted lines highlight the signal and sideband regions.} 
            \label{fig:BMass}
        \end{figure}

        In order to obtain the JFFs for \Bpm hadrons in jets, the distributions are measured for both signal and sideband regions of the \Bpm invariant mass, as shown in Fig.~\ref{fig:BMass}, and a sideband-subtraction procedure is used.  The signal region is defined in terms of 15 different \ptB bins from zero to $100\gevc$ based on the outcome of the fit to the \decay{\Bpm} {\jpsi \decay{(}{\mup \mun)} \Kpm} candidate invariant mass spectrum. For this procedure, the \Bpm-candidate mass is calculated with the constraint that the $\Bpm$ candidate must originate from the PV, with the dimuon invariant mass fixed to the known value of the \jpsi mass~\cite{PDG2024}. The asymmetric mass window around the known \Bpm mass value, along with these constraints makes the partially reconstructed background from \decay{\Bpm}{\jpsi (K / \pi) X} decays (where $X$ represents missed decay products) in the low mass tail negligible. The contribution from misreconstructed \decay{\Bpm}{\jpsi \pipm} decays, which are already suppressed in the branching fraction by a factor of 3.8\% with respect to the \decay{\Bpm}{\jpsi \Kpm} decay, are further reduced to a negligible level by the requirement on the kaon~PID.

        The signal probability density function (PDF) is defined as the sum of two double-sided Crystal Ball (DSCB) functions~\cite{Skwarnicki:1986xj} with the same peak position and tail parameters but independent widths. The tail parameters are fixed from simulation, while the peak position and width parameters remain free. The combinatorial background is modeled by a second-order Chebyshev polynomial. The level of background under the signal peak decreases as \ptB increases. The signal region is defined as $[\mu - N \sigma, \mu + N \sigma]$, where $\mu$ is the shared mean of the two DSCB functions, $\sigma$ is the width from the DSCB containing the majority of the signal after the fit, and $N = 2$ for $\ptB < 5 \gevc$, $N = 3$ for $5 < \ptB < 8\gevc$, and $N=4$ for $\ptB > 8 \gevc$. The width of the signal region is increased at higher \ptB to improve the signal selection. A representative mass fit within $8 < \ptB < 9 \gevc$ is shown in Fig.~\ref{fig:BMass}, with the signal and sideband regions highlighted by the green dotted and purple dash-dotted lines, respectively.

        As noted above, to account for combinatorial background present in the signal region, a sideband subtraction procedure is implemented for observables of interest. The observables are calculated with imposed requirements on the $\Bpm$ candidate mass to fall within the signal or sideband regions. The distribution from the sidebands is scaled by the ratio of integrals of the background PDF within the signal region to that within the sidebands, and then subtracted from the distribution calculated in the signal region. This statistically removes fake contributions to the observable distributions associated to the combinatorial background, assuming that the behavior of the combinatorial background is the same in the sidebands and in the signal region.

    \subsection{Corrections for detector effects}
    \label{sec:corrections}

        Due to the finite resolution of the LHCb detector and the composite nature of jets, several corrections must be applied that can impact the shape of the JFFs. A purity correction is applied to account for fake \B-jets entering the sample, for example those mapped to a true jet outside of the reported kinematic region. An unfolding procedure is applied to correct for the detector response in the measured variables. Finally, an efficiency correction is applied to account for missed \B-jets. These corrections are derived by comparing simulation samples at truth and reconstructed level, with further data-driven corrections applied to factorized components of the overall efficiency. 

        A geometric matching procedure is applied to simulated \B-jets at truth and reconstructed level, from which the purity, detector response, and efficiency are measured. Jets at reconstructed level must have a corresponding $B$-tagged jet at truth level within $\Delta R < 0.5$. The purity is measured as the ratio of the number of matched \B-jets to the total number of \B-jets at reconstructed level in bins of the respective observable distributions. This provides a measure of the probability that an observed \B-jet is a signal jet (or truly produced within LHCb acceptance). The purity corrections are applied as a bin-by-bin multiplicative scale factor.

        The detector response is quantified via response matrices in the variables of interest for matched candidates between reconstructed and truth level simulation. An iterative Bayesian unfolding procedure is applied using the \textsc{RooUnfold} package to correct the measured distributions for smearing due to detector effects~\cite{ unfolding_bayesian,RooUnfold}. The \ptj-binned collinear JFFs and additional 1D distributions have a 4D response matrix (RM) in $(z, \ptj), (\jt, \ptj)$, and $(r, \ptj)$ at truth and reconstructed levels respectively, while the \ptj-binned TMD JFFs and additional 2D correlations with radial profile $r$ have a 6D RM in $(z, \jt, \ptj), (z, r, \ptj)$, and $(\jt, r, \ptj)$ at truth and reconstructed levels respectively. The prior distribution used for the unfolding is adjusted at each successive iteration, starting with the truth-level projection of the RM based on the \mbox{\pythia8} event generator with an LHCb tune~\cite{Sjostrand:2007gs, LHCb-PROC-2010-056}. Given that JFFs and related observables are sensitive to the hadronization process, the hadronization model in the simulation sample can bias the correction scheme. To account for this, variations on the shape of the prior distribution are used to unfold the data, from which systematic uncertainties are derived (discussed in Sec.~\ref{sec:sys}).

        After the data are unfolded, the distributions are corrected for the efficiency of reconstructing a \B-jet given the selection criteria. This is defined as the ratio of matched \B-jets to the number produced in the truth-level simulation. This provides a measure of the probability that a \B-jet is reconstructed and subsequently selected given that one was truly produced, and quantifies the fraction of missed \B-jets. The efficiency corrections are applied by dividing the observable distributions by the efficiency in each bin.

        The total \B-jet reconstruction efficiency, aside from the PID efficiency, is calculated using simulation and adjusted with data-driven methods. This embodies the tracking efficiency, trigger efficiency, as well as the reconstruction and selection efficiency of the \B-jet, and varies between ${\sim}10 \% - 25 \%$.
        Adjustments are made to the tracking and trigger efficiency by using standard LHCb tools~\cite{LHCb-DP-2013-002, LHCb-PUB-2014-039}, through which each entry of the observable histograms is weighted by the inverse ratio of these efficiencies in data to those in simulation. These ratios for track efficiency are applied for the \Bpm decay products, and vary from ${\sim}0.95 - 1.03$ in most track $(p, \eta)$ bins, dropping to ${\sim}0.85$ at lower track momentum. The trigger efficiency ratios are applied in bins of \jpsi transverse momentum and rapidity and vary from ${\sim}1.05 - 1.2$. The efficiency of the PID selection criteria is instead measured in a purely data-driven manner, and the corresponding corrections are applied by ensuring that each entry of the observable histograms is weighted by the inverse of the PID efficiency measured on a calibration sample from data~\cite{LHCb-PUB-2016-021}. The PID efficiencies are measured separately for muons and kaons. The muon PID efficiencies are typically above $0.9$ and increase with $p$, while suffering from edge effects at higher $\eta$ leading to efficiencies varying between ${\sim}0.8 - 0.9$. The kaon PID efficiencies are typically above $0.95$ at low kaon momentum, dropping to ${\sim}0.6$ as $p$ increases, and experiencing a similar ${\sim} 10 \%$ degradation of the efficiencies at high $\eta$. These corrections are performed on an event-by-event basis given that the tracking, PID, and trigger efficiencies depend on event-specific kinematic variables such as particle momentum and pseudorapidity.

\section{Systematic uncertainties}
\label{sec:sys}

    Various sources of systematic uncertainty, related to the selection, the signal extraction, and corrections for detector effects, are investigated. A variation is performed on the $\Bpm$ reconstruction where requirements to ensure a good-quality decay vertex that is well separated from the PV are tightened for both the $\Bpm$ and $\jpsi$ candidates. A modified jet-selection criterion is also tested, where at least one track must be reconstructed in addition to the $\Bpm$ meson. In both cases, the new selection criteria are applied to both data and simulation, and the analysis chain is repeated, with the systematic uncertainty calculated as the difference between the baseline results and those obtained from the systematic variation. For the signal extraction, variations are performed on the fit model, where the signal model is replaced by a Student's t-distribution~\cite{Student_T}, and the definition of the sideband regions used in the combinatorial background subtraction procedure. Two different definitions for the sideband regions are tested and the uncertainties obtained from these variations are averaged before combining with the rest of the sources of uncertainty. The systematic variations on the signal extraction procedure are only applied to data. 
    
     Additional sources of systematic uncertainty come from the procedure to correct the distributions for detector effects. These are further categorized into sources related to the data-driven efficiency corrections, the jet energy scale and resolution (JES and JER), and the unfolding procedure. Efficiencies related to PID, tracking, and triggering for the $\Bpm$ decay products are corrected for in a data-driven manner (see Sec.~\ref{sec:corrections}). For the muon tracking efficiencies, as well as PID efficiencies, the statistical uncertainties on the calibration sample are propagated by shifting the event weights by $\pm 1 \sigma$ and recalculating the final distributions, while for kaon candidates, an additional uncertainty from hadronic interactions is propagated to the tracking efficiency calculation~\cite{LHCb-DP-2013-002}. The data-driven method for determining the trigger efficiency is compared with an estimate obtained purely from simulation, and the deviation from unity in the ratios obtained between the two methods is taken as an uncertainty on the trigger efficiency. The event weights for the trigger efficiency ratio are shifted by this uncertainty and the final distributions are recalculated. In both cases, the uncertainty is propagated to the final results by taking differences between the baseline distributions and those obtained from the respective systematic variations. The resulting $\pm 1 \sigma$ deviations are averaged in each case to obtain a single systematic uncertainty for each source.

    A systematic uncertainty for the jet energy scale and resolution is obtained by studying jets recoiling from a \decay{\Z}{\mu^{+} \mu^{-}} decay. Selection criteria are applied to ensure clean back-to-back events are observed. The jets must satisfy $|\Delta \phi| > \frac{7\pi}{8}$ with respect to the \Z boson, and the subleading jet must have a $p_{\rm T,\text{subleading}} < 0.25 \cdot p_{\rm T,\text{leading}}$, where the subleading jet is defined as the jet with the second-largest $\ptj$ in the event. The \pt balance $\ptj / p_{\rm T, Z}$ is measured in both data and reconstructed-level simulation, and parameters to shift and smear the \pt balance distribution at reconstructed level to best match that of data are determined simultaneously. The corresponding shift and smear factors, related to the JES and JER, respectively, are determined in various $\ptj$ bins, and correspond to about a $-3 \%$ shift and $12 \%$ smearing on average. Jets in the reconstructed-level simulation sample used for the $\Bpm$ JFF analysis are then adjusted according to the JES and JER values, leading to a new set of detector corrections. Given that the jet momentum enters directly into the calculation of the observables [Eqs.~(\ref{eq:frag_z})--(\ref{eq:frag_r})], this is one of the dominant systematic uncertainties, particularly at low \ptj where the JES shift is larger. 

    The uncertainty of the remaining detector corrections is estimated through a series of closure tests, as well as by varying the number of iterations during the unfolding procedure and the prior distribution used to unfold the data. The closure tests are conducted by applying the detector correction chain on observable distributions from a simulated sample and testing if the truth distributions are reproduced. Both a statistical and a shape closure test are performed to demonstrate the robustness of the correction procedure to statistical fluctuations in the measured distribution and differing shapes between observables in data and reconstructed-level simulation. For the statistical closure test, the reconstructed-level simulation is fluctuated according to the statistical uncertainties in the corresponding distribution in data, and the full set of detector corrections is applied to the modified sample. Projections in various \ptj bins are then normalized and compared to the normalized distributions at truth level by taking ratios. The procedure is repeated 10 times with a different random seed, and the bin-by-bin differences from unity are averaged in quadrature to quantify the degree of nonclosure. The shape closure test is conducted on the baseline reconstructed-level simulation (\ie not fluctuated according to uncertainties in data), but with an RM that is reweighted by the ratio of the $(z, \jt, \ptj)$ distribution in data (after the sideband subtraction procedure) to that in reconstructed-level simulation. The correction scheme is applied to the reconstructed-level simulation distributions, but with the modified RM, and ratios of the normalized projections in various \ptj bins to the truth-level simulation are taken. In both closure tests, the ratios of corrected reconstructed-level simulation to truth-level simulation have corresponding statistical uncertainties. A systematic uncertainty is only assigned for nonclosure if the ratios are more than $1 \sigma$ statistical uncertainty away from unity by subtracting the statistical uncertainty from the central value in quadrature. A systematic variation of the number of unfolding iterations ($\pm 1$ iteration) is also performed, where resulting deviations from the baseline results are averaged for the two cases. Finally, a systematic variation of the prior distribution used to unfold the data is obtained by again reweighting the RM by the $(z, \jt, \ptj)$ distribution in data to that in reconstructed-level simulation. Given that these four sources of systematic uncertainty are all probing the underlying uncertainty on the unfolding procedure and corrections for detector effects, the correlated sources are averaged before combining with other sources of systematic uncertainty. 

    The total systematic uncertainty is obtained by treating the distinct sources as independent and summing them in quadrature. The unfolding and JES/JER systematic uncertainties are dominant, with the JES/JER having a larger impact at lower \ptj, and the unfolding uncertainties having a larger impact at higher \ptj. Systematic uncertainties related to event and candidate selection, signal extraction, and efficiency corrections are about an order of magnitude smaller than the dominant systematic uncertainties in most bins.

\section{Results and discussion}
\label{sec:results}
\begin{figure}[H]		
        \centering
        \includegraphics[width=\textwidth,height=\textheight, keepaspectratio]{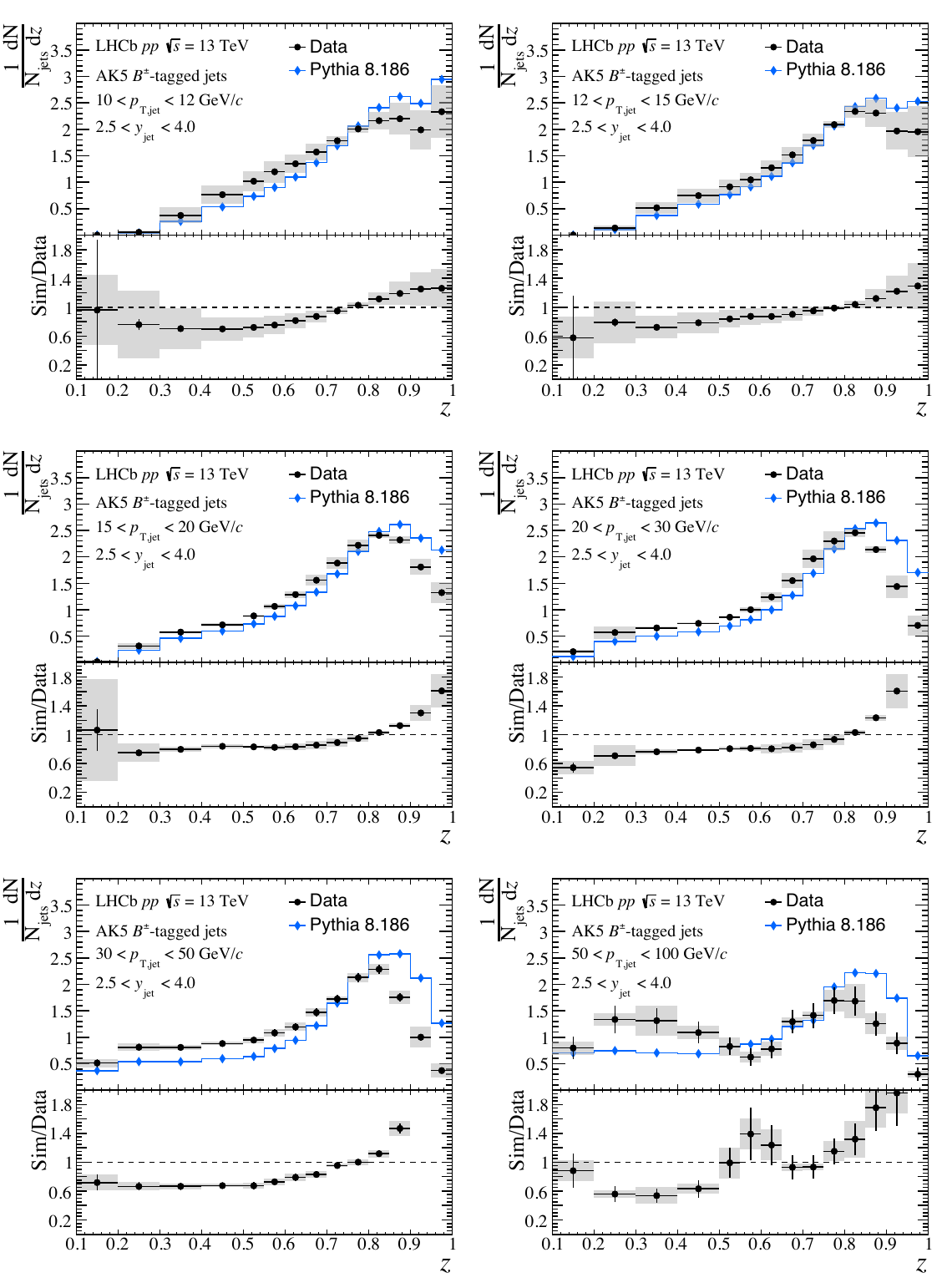}
        \caption{Measured $z$ distributions (top panels) shown alongside simulation/data ratios (bottom panels) in each \ptj bin. Note that some points are outside of the displayed range in the bottom panel. In the top panels, the black points and bars correspond to the data points and statistical error bars, the shaded gray boxes correspond to the systematic uncertainties, and the blue points correspond to the prediction from \pythia 8.186. The measured quantity corresponds to the collinear JFF for \Bpm mesons.}
        \label{fig:publish_z_all}
    \end{figure}

\begin{figure}[H]		
        \centering
        \includegraphics[width=\textwidth,height=\textheight, keepaspectratio]{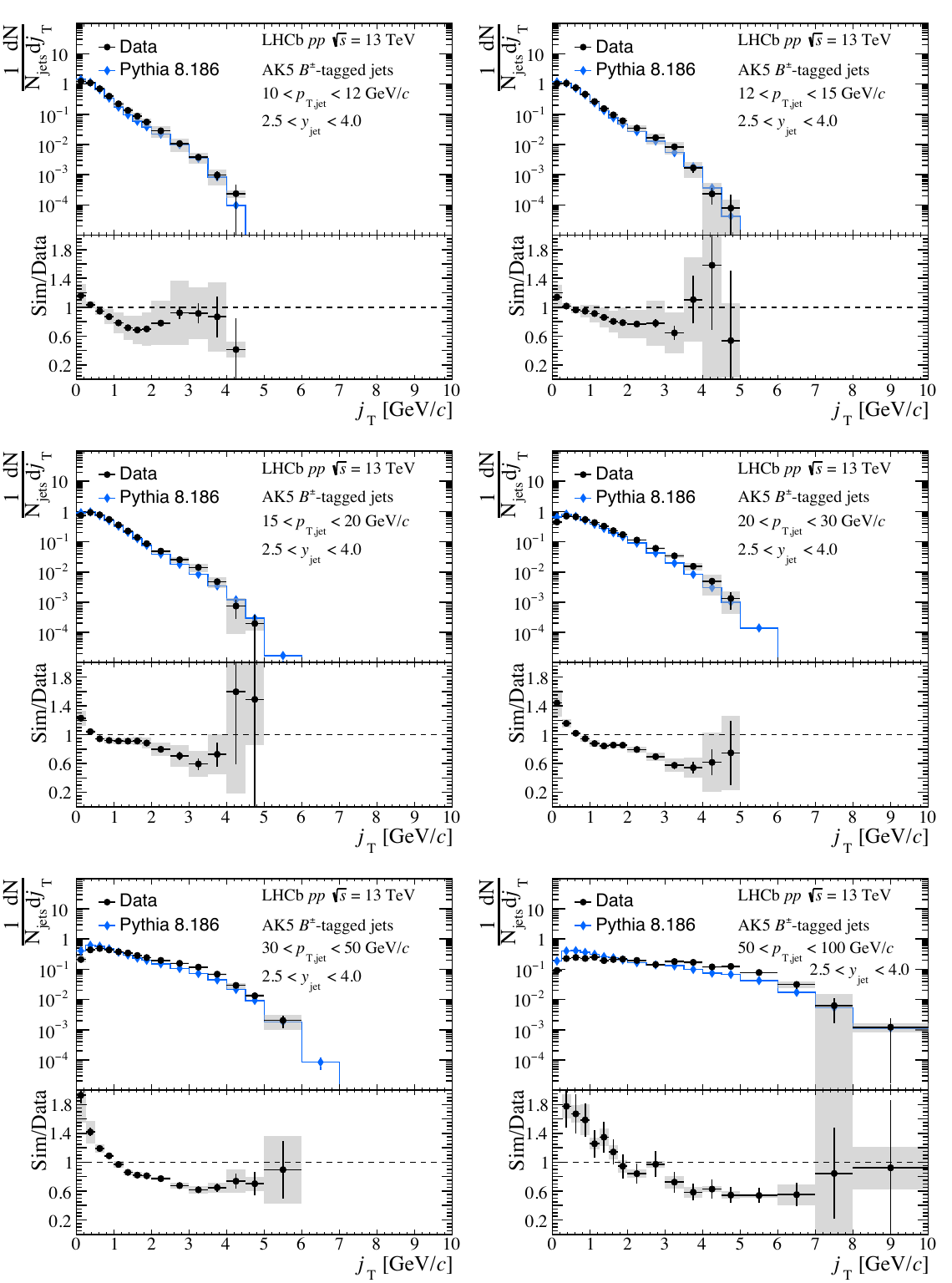}
        \caption{Measured \jt distributions (top panels) shown alongside simulation/data ratios (bottom panels) in each \ptj bin. In the top panels, the black points and bars correspond to the data points and statistical error bars, the shaded gray boxes correspond to the systematic uncertainties, and the blue points correspond to the prediction from \pythia 8.186.}
        \label{fig:publish_jt_all}
    \end{figure}

\begin{figure}[H]		
        \centering
        \includegraphics[width=\textwidth,height=\textheight, keepaspectratio]{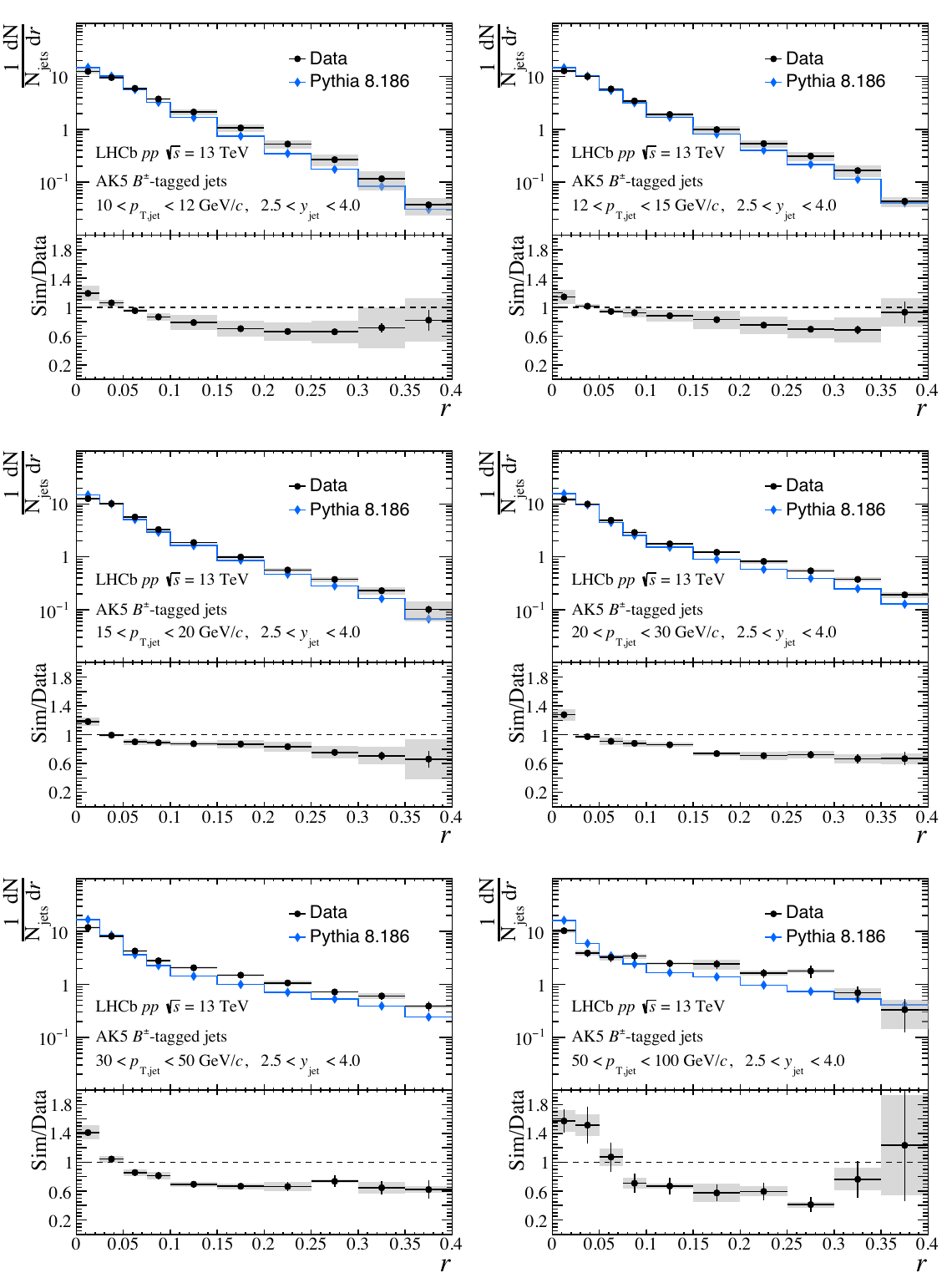}
        \caption{Measured $r$ distributions (top panels) shown alongside simulation/data ratios (bottom panels) in each \ptj bin. In the top panels, the black points and bars correspond to the data points and statistical error bars, the shaded gray boxes correspond to the systematic uncertainties, and the blue points correspond to the prediction from \pythia 8.186.}
        \label{fig:publish_r_all}
    \end{figure}

\begin{figure}[H]		
        \centering
        \includegraphics[width=\textwidth,height=\textheight, keepaspectratio]{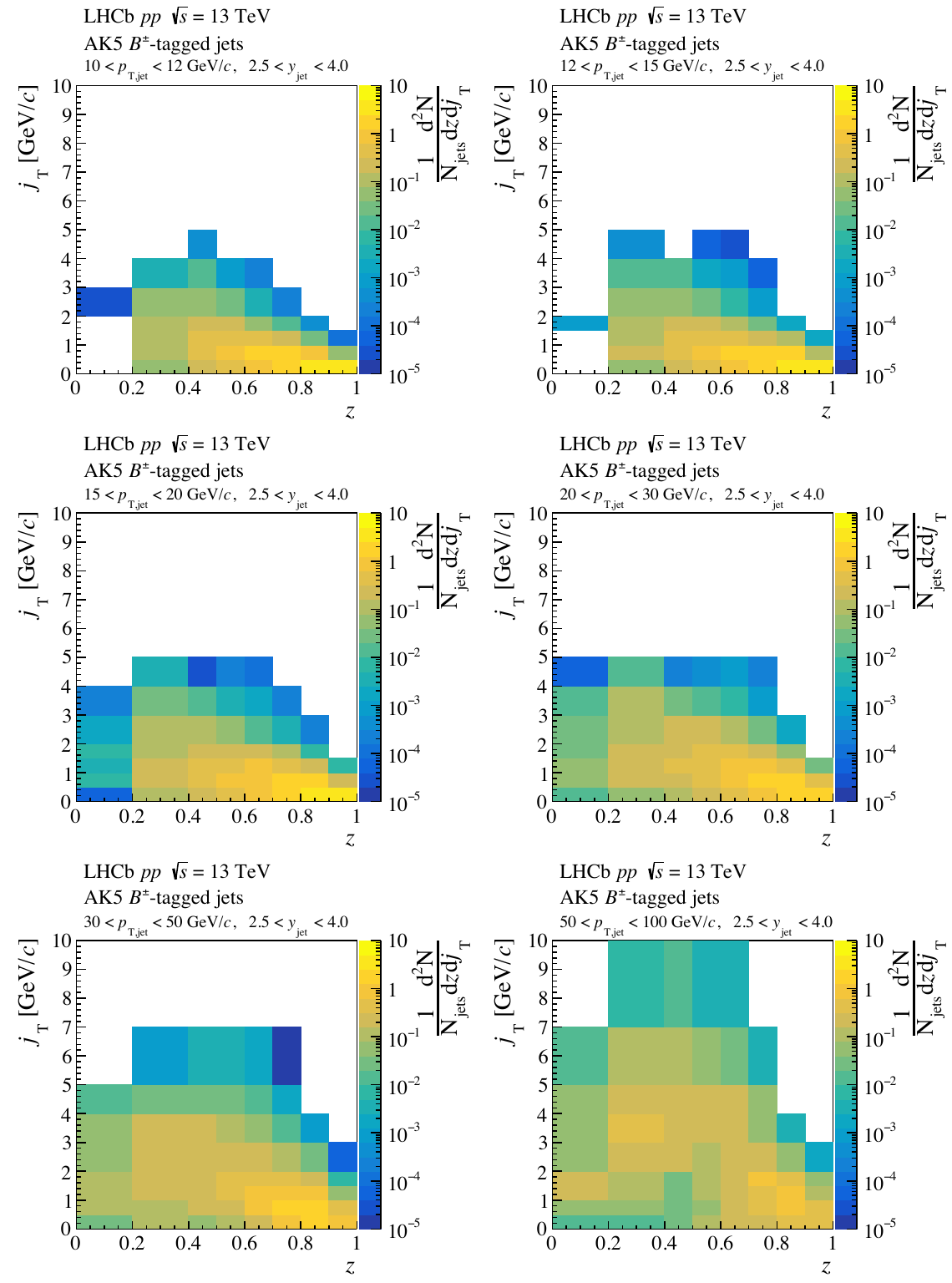}
        \caption{Measured $(z, \jt)$ distributions in each \ptj bin. The measured quantity corresponds to the TMD JFF for \Bpm mesons. A growing contribution at low $z$ and high \jt can be observed with increasing \ptj.}
        \label{fig:publish_zjt_all}
    \end{figure}

\begin{figure}[H]		
        \centering
        \includegraphics[width=\textwidth,height=\textheight, keepaspectratio]{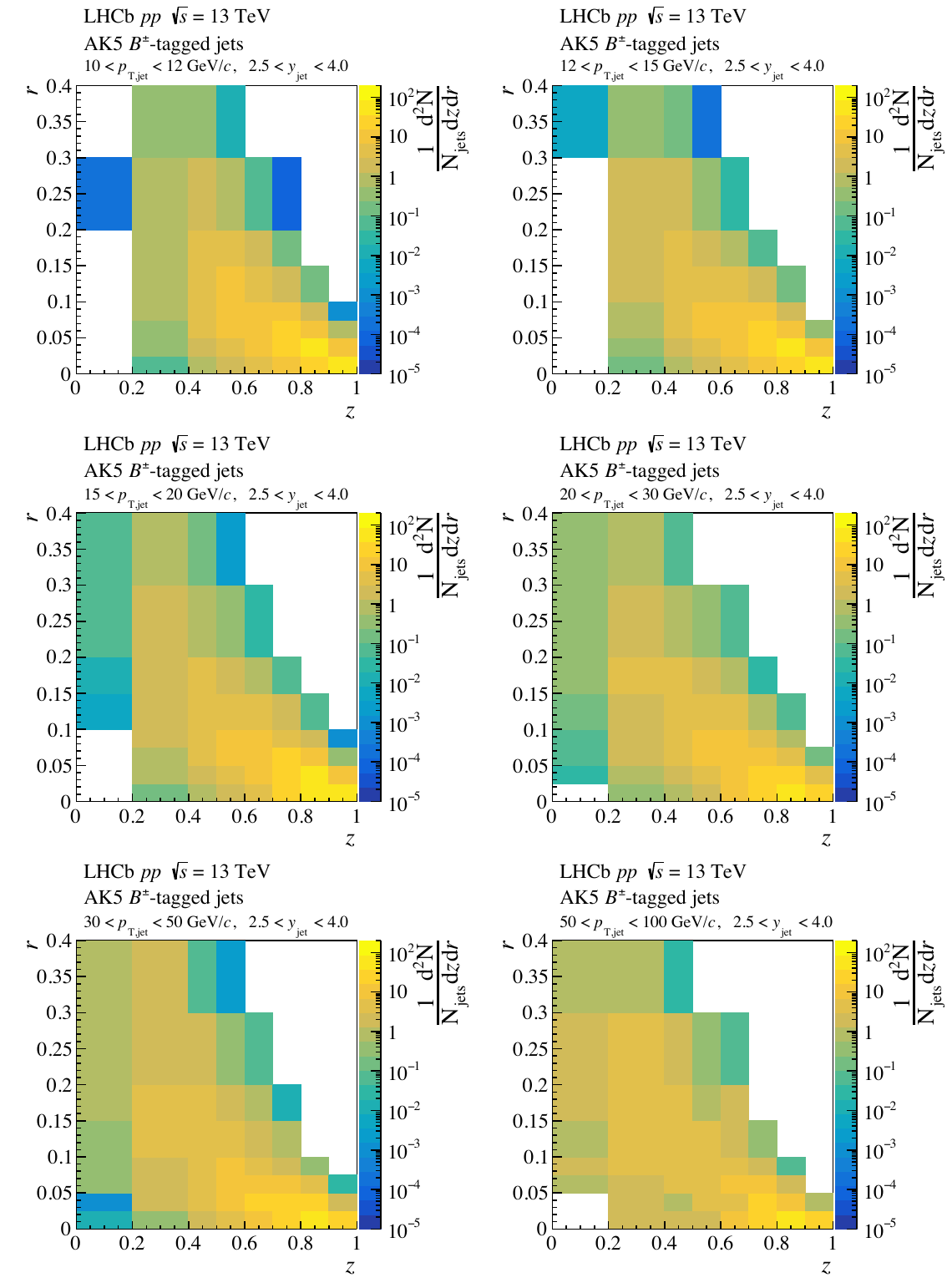}
        \caption{Measured $(z, r)$ distributions in each \ptj bin. A growing contribution at low $z$ and high $r$ can be observed with increasing \ptj.}
        \label{fig:publish_zr_all}
    \end{figure}

\begin{figure}[H]		
        \centering
        \includegraphics[width=\textwidth,height=\textheight, keepaspectratio]{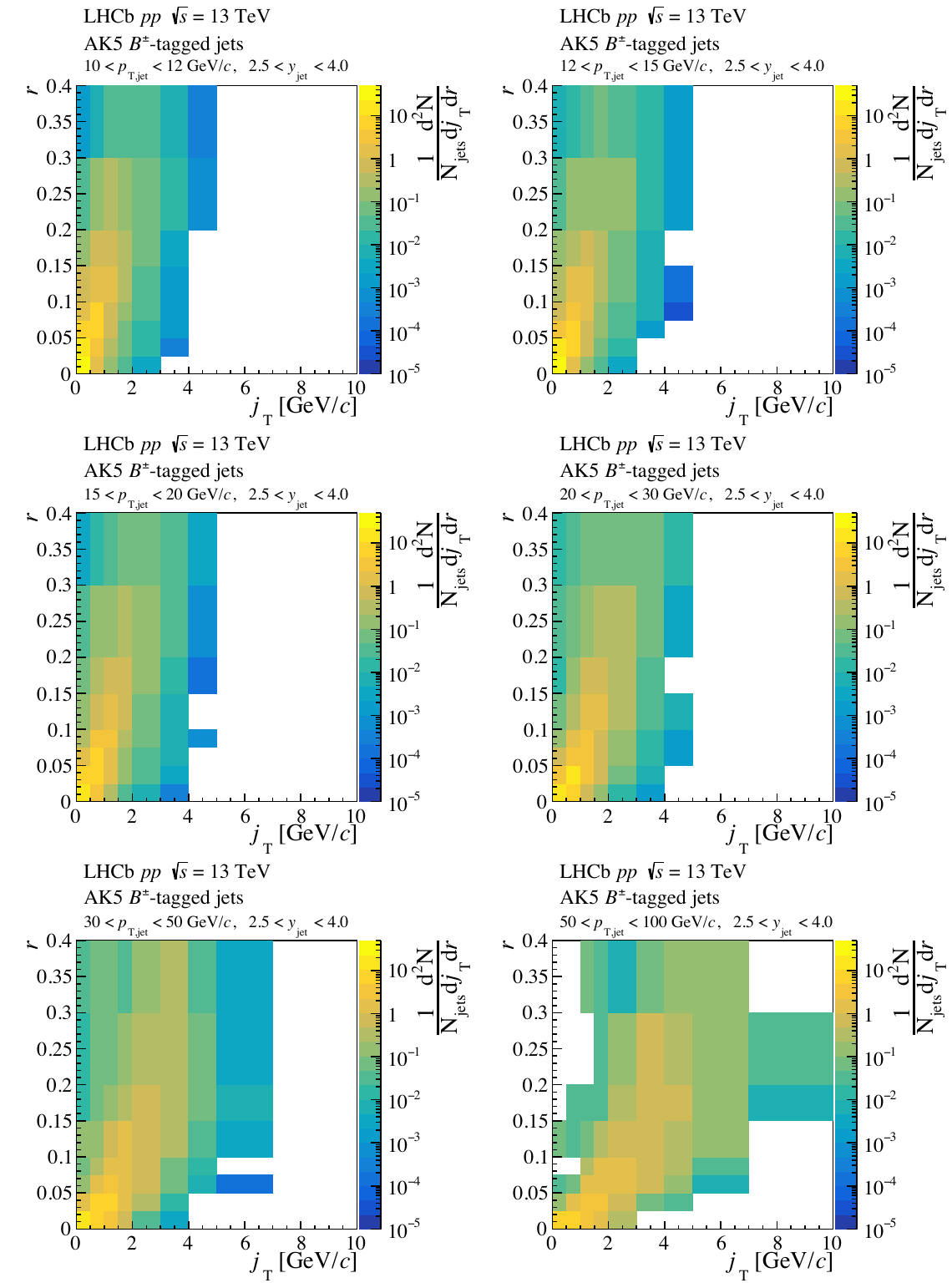}
        \caption{Measured $(\jt, r)$ distributions in each \ptj bin. A growing contribution at high \jt and high $r$ can be observed with increasing \ptj.}
        \label{fig:publish_jtr_all}
    \end{figure}

    The resulting distributions for $z, \jt,$ and $r$ as well as $(z, \jt), (z, r),$ and $(\jt, r)$ are shown in Figs.~\ref{fig:publish_z_all}--\ref{fig:publish_jtr_all} in six \ptj bins ranging from 10 to 100\gevc. The distributions are normalized by the number of jets in the corresponding \ptj bin, with the final results taking the form
    
    \begin{equation}
        \label{eq:final_observables}
        \frac{1}{N_{\rm jets}}\frac{\deriv N}{\deriv \mathcal{O}},
    \end{equation}
    where $\deriv\mathcal{O}$ is replaced by $\deriv z, \deriv\jt,$ or $\deriv r$ for the 1D distributions and by $\deriv z \deriv \jt$, $\deriv \jt \deriv r$, or $\deriv z\deriv r$ for the joint distributions. In the case of $\deriv\mathcal{O} = \deriv z$ and $\deriv \mathcal{O} = \deriv z\deriv \jt$, Eq.~(\ref{eq:final_observables}) reproduces the collinear, $F(z)$, and TMD, $F(z, \jt)$, JFFs, respectively. These measurements include the same observable measured by the ATLAS experiment~\cite{ATLAS_Bfrag}, with extended reach in the low \ptj region down to $10\gevc$ (compared to $50\gevc$ from the ATLAS measurement), and in a complementary rapidity region of $2.5 < y_{\rm jet} < 4.0$ (compared to $| \eta_{\rm jet} | < 2.1$). All of the results reported in this paper, including the statistical and systematic uncertainties, are provided in a \textsc{HEPData} record~\cite{HEPData_entry}.

    Figure~\ref{fig:publish_z_all} shows the final $z$ distribution in each \ptj bin, which corresponds to the collinear JFF $F(z)$. The data are plotted alongside truth-level simulation obtained from \pythia 8.186 in the top panel of each plot, while the ratio of the \pythia prediction to data is shown in the bottom panel of each plot. The data show an increase in the JFF at low $z$ as \ptj increases, which is understood to be due to a growing contribution of gluon fragmentation $g \to \Bpm$ with an intermediate $g \to b \bar{b}$ splitting that reduces the collinear momentum fraction of the observed $\Bpm$ meson. It can also be seen, in particular for increasing \ptj, that \pythia tends to predict larger isolated production of $\Bpm$ mesons and thus likely predicts lower gluon-splitting contributions. Furthermore, the forward-rapidity coverage of the LHCb detector provides a unique phase space that should result in an increased gluon-splitting fraction relative to midrapidity detectors for a fixed \ptj interval, making these data particularly valuable together with the collinear JFF measurement for \Bpm mesons by the ATLAS Collaboration~\cite{ATLAS_Bfrag}. A larger prediction of isolated heavy-flavor production by \pythia was also observed for prompt \jpsi mesons~\cite{LHCB-PAPER-2016-064}, as well as prompt $\psi(2S)$ and exotic $\chicone(3872)$ mesons~\cite{LHCb-PAPER-2024-021}. Figures~\ref{fig:publish_jt_all} and~\ref{fig:publish_r_all} show the results for \jt and $r$, respectively, in increasing \ptj bins. Figure~\ref{fig:publish_jt_all} shows the broadening of the \jt distributions with increasing \ptj, partly due to the increase in kinematic reach, but also potentially due to a growing contribution from gluon fragmentation to $\Bpm$, where the $g \to b\bar{b}$ splitting would lead to some transverse momentum of the $b$ and $\bar{b}$ relative to the outgoing gluon. Figure~\ref{fig:publish_r_all} shows that the $\Bpm$ is typically quite close to the jet axis, which makes intuitive sense given the high collinear momentum fractions shown in Fig.~\ref{fig:publish_z_all}. The radial profile also tends to broaden out as \ptj increases, potentially due to the increased contribution of gluon fragmentation. It can be seen again in the ratios of Fig.~\ref{fig:publish_r_all} that \pythia tends to predict larger isolated $\Bpm$ production. This may suggest a more significant contribution of in-shower production of heavy-flavor quarks.

    Figure~\ref{fig:publish_zjt_all} shows the data distributions for $(z, \jt)$, corresponding to the first measurement of the TMD JFF $F(z, \jt)$ for $\Bpm$ mesons. The anticorrelation between $z$ and \jt can be observed for $\Bpm$ mesons in jets, with higher $z$ $\Bpm$ mesons typically carrying a lower \jt. Similarly, a strong anticorrelation between $(z,r)$ is shown in Fig.~\ref{fig:publish_zr_all} as one would expect. Figure~\ref{fig:publish_jtr_all} shows the results for $(\jt, r)$ where instead a strong correlation can be observed, as $\Bpm$ mesons with larger \pt relative to the jet axis are more likely to have a larger angular separation from the jet axis. Across the distributions, the systematic uncertainties vary from being roughly an order of magnitude larger than the statistical uncertainties in the lower \ptj bins to being comparable to the statistical uncertainties in the higher \ptj bins.

\section{Summary}
\label{sec:summary}

    The transverse momentum dependent and collinear jet fragmentation functions, as well as the radial profile with respect to the jet axis have been measured for $\Bpm$ hadrons in jets in $\sqs = 13\tev$ $pp$ collisions by the LHCb Collaboration. The $\Bpm$ mesons are reconstructed through the $\jpsi (\to \mup \mun) \Kpm$ decay channel and clustered into anti-$k_{t}$ jets with a resolution parameter of $R = 0.5$. This measurement provides useful input to global analyses of $\Bpm$ fragmentation, for which the collinear jet fragmentation function has also been measured by ATLAS~\cite{ATLAS_Bfrag} in a complementary \ptj and $\eta_{\rm jet}$ region, in particular for better constraining the contribution from gluon-fragmentation process $g \to \Bpm$. The results for the collinear jet fragmentation function as well as the radial profile show that \pythia tends to overpredict isolated production of $\Bpm$ mesons, in particular as \ptj increases, suggesting a scale-dependent contribution from heavy-flavor production in the parton shower. Further investigation and comparison of JFFs for different beauty and charm mesons and baryons as a function of \ptj will help shed light on contributions from in-shower heavy-flavor production.  In addition, the TMD JFF as well as the other joint distributions $(z, r)$ and $(\jt, r)$ for $\Bpm$ mesons are reported for the first time. This complements a growing body of work for jet fragmentation functions of different hadronic species, allowing for an understanding of the mass dependence of the fragmentation process. In addition, together with recent measurements of JFFs for charged particles in heavy-flavor tagged jets~\cite{LHCb-PAPER-2025-038}, the Lund Jet Plane for \Bpm jets~\cite{LHCb-PAPER-2025-010}, and the jet mass for \Bpm jets~\cite{LHCb-PAPER-2025-009}, a more complete picture of the parton-to-jet transition for heavy-flavor jets emerges. 

   % z %
    % jt %
    % r %
    % (z, jt)
    % (z, r)
    % (jt, r)
 
%refs

% Do not include this in any draft (just for information in the template)
%\input{LHCbInternal/acknowledgements_template}
% Comment this in for paper drafts; do not include this in analysis note, conference and figure reports
%\input{LHCb/acknowledgements}

\section*{Acknowledgements}
%
% These Acknowledgements valid from 21/02/26
%
\noindent We express our gratitude to our colleagues in the CERN
accelerator departments for the excellent performance of the LHC. We
thank the technical and administrative staff at the LHCb
institutes.
We acknowledge support from CERN and from the national agencies:
ARC (Australia);
CAPES, CNPq, FAPERJ and FINEP (Brazil); 
MOST and NSFC (China); 
CNRS/IN2P3 and CEA (France);  % added CEA 26/02/2026
BMFTR, DFG and MPG (Germany);
INFN (Italy); 
NWO (Netherlands); 
MNiSW and NCN (Poland); 
MCID/IFA (Romania); 
%MSHE (Russia); 
MICIU and AEI (Spain);
SNSF and SER (Switzerland); 
NASU (Ukraine); 
STFC (United Kingdom); 
DOE NP and NSF (USA).
%%%%%%%%%%%%%%%%%%%%%%%%%%%%%%%%%%%%%%%%%%%%%
We acknowledge the computing resources that are provided by ARDC (Australia), 
CBPF (Brazil),
CERN, 
IHEP and LZU (China),
IN2P3 (France), 
KIT and DESY (Germany), 
INFN (Italy), 
SURF (Netherlands),
Polish WLCG (Poland),
IFIN-HH (Romania), 
%RRCKI and Yandex LLC (Russia), 
PIC (Spain), CSCS (Switzerland), 
GridPP (United Kingdom),
and NSF (USA).  % added Feb2026
%%%%%%%%%%%%%%%%%%%%%%%%%%%%%%%%%%%%%%%%%% 
We are indebted to the communities behind the multiple open-source
software packages on which we depend.
%%%%%%%%%%%%%%%%%%%%%%%%%%%%%%%%%%%%%%%%%%
Individual groups or members have received support from
% ARC and ARDC (Australia); % moved to national 16/01/2025
RTP (Australia), % added 06/03/2026
Key Research Program of Frontier Sciences of CAS, CAS PIFI, CAS CCEPP (China); 
%Fundamental Research Funds for the Central Universities,  and Sci.\
%\& Tech.\ Program of Guangzhou (China); Removed 24/11/25
Minciencias (Colombia);
EPLANET, Marie Sk\l{}odowska-Curie Actions, ERC and NextGenerationEU (European Union);
A*MIDEX, ANR, IPhU and Labex P2IO, and R\'{e}gion Auvergne-Rh\^{o}ne-Alpes (France);
%RFBR, RSF and Yandex LLC (Russia);
Alexander-von-Humboldt Foundation (Germany);
ICSC (Italy); 
%GVA, XuntaGal, GENCAT, Inditex, InTalent and Prog.~Atracci\'on Talento, CM (Spain);
Severo Ochoa and Mar\'ia de Maeztu Units of Excellence, GVA, XuntaGal, GENCAT, InTalent-Inditex and Prog.~Atracci\'on Talento CM (Spain);
%XuntaGal --> Xunta de Galicia 
% SRC (Sweden);  % removed 27/02/2026 - end of grant
the Leverhulme Trust, the Royal Society and UKRI (United Kingdom).

\section*{Data availability}
The data that support the findings of this article are openly available~\cite{HEPData_entry}.

%\input{LHCbInternal/supplementary_template}

%\input{appendix}

% This should be taken out in the final paper
%\input{supplementary}

\addcontentsline{toc}{section}{References}
%\setboolean{inbibliography}{true}
%\bibliographystyle{LHCb/LHCb}
%\bibliography{main,LHCb/standard,LHCb/LHCb-PAPER,LHCb/LHCb-CONF,LHCb/LHCb-DP,LHCb/LHCb-TDR}
%\input{main.bbl}
\ifx\mcitethebibliography\mciteundefinedmacro
\PackageError{LHCb.bst}{mciteplus.sty has not been loaded}
{This bibstyle requires the use of the mciteplus package.}\fi
\providecommand{\href}[2]{#2}

\newpage
%\input{Authorship_LHCb-PAPER-2025-061}
%\input{~/Downloads/Authorship_LHCb-SD-2024-001-grouped-internal-use-only-do-not-publish-2.tex}
% LHCb collaboration author list
% Data extracted on March 18th, 2026 at 2:44pm for paper reference LHCb-PAPER-2025-061
\centerline
{\large\bf LHCb collaboration}
\begin
{flushleft}
\small
R.~Aaij$^{38}$\lhcborcid{0000-0003-0533-1952},
A.S.W.~Abdelmotteleb$^{58}$\lhcborcid{0000-0001-7905-0542},
C.~Abellan~Beteta$^{52}$\lhcborcid{0009-0009-0869-6798},
F.~Abudin\'en$^{60}$\lhcborcid{0000-0002-6737-3528},
T.~Ackernley$^{62}$\lhcborcid{0000-0002-5951-3498},
A. A. ~Adefisoye$^{70}$\lhcborcid{0000-0003-2448-1550},
B.~Adeva$^{48}$\lhcborcid{0000-0001-9756-3712},
M.~Adinolfi$^{56}$\lhcborcid{0000-0002-1326-1264},
P.~Adlarson$^{88}$\lhcborcid{0000-0001-6280-3851},
C.~Agapopoulou$^{14}$\lhcborcid{0000-0002-2368-0147}
% ...

{\footnotesize \it

% ...
$^{14}$Universit{\'e} Paris-Saclay, CNRS/IN2P3, IJCLab, Orsay, France\\
% ...
$^{38}$Nikhef National Institute for Subatomic Physics, Amsterdam, Netherlands\\
% ...
$^{48}$Instituto Galego de F{\'\i}sica de Altas Enerx{\'\i}as (IGFAE), Universidade de Santiago de Compostela, Santiago de Compostela, Spain\\
% ...
$^{52}$Physik-Institut, Universit{\"a}t Z{\"u}rich, Z{\"u}rich, Switzerland\\
% ...
$^{56}$H.H. Wills Physics Laboratory, University of Bristol, Bristol, United Kingdom\\
% ...
$^{58}$Department of Physics, University of Warwick, Coventry, United Kingdom\\
% ...
$^{60}$School of Physics and Astronomy, University of Edinburgh, Edinburgh, United Kingdom\\
$^{61}$School of Physics and Astronomy, University of Glasgow, Glasgow, United Kingdom\\
$^{62}$Oliver Lodge Laboratory, University of Liverpool, Liverpool, United Kingdom\\
% ...
$^{70}$Syracuse University, Syracuse, NY, United States\\
% ...
$^{88}$Department of Physics and Astronomy, Uppsala University, Uppsala, Sweden, associated to $^{61}$\\
% ...
}
\end{flushleft}

\end{document}